# Antimicrobial resistance and use, and rates of hospitalization associated with bacterial infections, including sepsis

Edward Goldstein[1,*], Derek R. MacFadden[1], Marc Lipsitch[1,2]

1. Center for Communicable Disease Dynamics, Department of Epidemiology, Harvard TH Chan School of Public Health, Boston, MA 02115 USA
2. Department of Immunology and Infectious Diseases, Harvard TH Chan School of Public Health, Boston, MA 02115 USA
* Corresponding author, egoldste@hsph.harvard.edu

## Abstract

While the mechanisms and quantitative details are complex, few analysts would doubt that antibiotic use increases the prevalence of drug-resistant bacterial pathogens among all bacteria causing disease in a population. The causal connection between antibiotic use and the total incidence of severe bacterial infections, possibly mediated by antibiotic resistance, is less clearly established. The increasing burden of severe bacterial infections and their sequelae, particularly sepsis, in the United States and other countries, calls out for an explanation. In this Perspective we consider the evidence bearing on the hypothesis that prevalence of antibiotic resistance and levels of antibiotic use are important contributors to the rates of sepsis hospitalizations and other outcomes with bacterial etiology. In the process, we discuss the consequences of resistance to/use of commonly prescribed antibiotics, including fluoroquinolones, and provide a comparison of the epidemiology of antibiotic use/resistance and severe outcomes associated with infections with bacterial pathogens, particularly *Escherichia coli*, methicillin-resistant *Staphylococcus aureus* (MRSA), and *Clostridium Difficile* in the UK vs. US.

Rates of hospitalization with septicemia or sepsis, and related mortality, as well as the associated monetary costs have been increasing during the past decades in the US [1-3]. The estimated number of US hospitalizations with septicemia (ICD-9 codes 038.xx present on the discharge diagnosis) nearly doubled from 836,600 in 2000 to 1,665,500 in 2009 [1]. During the same period, the estimated number of in-hospital deaths for hospitalizations with septicemia increased by 61%, from 163,688 to 267,329 (Table 1 in [1]). The CDC estimates that about 250,000 Americans die from sepsis each year, with one in three patients who die in a hospital having sepsis [4]. These mortality counts underestimate the number of deaths resulting from hospitalizations with septicemia in the US. For US hospitalizations with septicemia/sepsis in the elderly, the 90-day case fatality rate is significantly higher than the in-hospital mortality rate [5]. Therefore, the number of sepsis-related deaths in the US would be substantially larger than the CDC estimate in [4] if 90-day mortality were considered. Also, only a fraction of septicemia-related deaths, even those that take place in the hospital, have septicemia listed on the death certificate ([1] vs. [6]). In fact, there is a significant discrepancy between hospital discharge diagnoses and the lists of causes of death on death certificates in the US, with death certificates potentially underestimating infections and sepsis as proximal causes of death [7,8].

The estimated number of hospitalizations with septicemia in the principal discharge diagnosis increased by a further 81% between 2009-2014 [1,2]. Hospitalizations with septicemia in the principal diagnosis alone in 2013 cost over $23 billion [3]. However, rates of hospitalizations meeting certain clinical criteria (defined in [9]) indicative of sepsis remained stable during the same period [9]. Moreover, the proportion of hospitalizations with septicemia/sepsis in the diagnosis that have a positive blood culture has declined in various places in the US [10]. Change in diagnostic practices for septicemia/sepsis is an important contributing factor to the increase in the volume of hospitalizations with septicemia/sepsis in the diagnosis, especially since mid 2000s. In particular, adoption of electronic sepsis screening and treatment protocols took place in many hospitals, with such protocols having both beneficial and adverse effects [11,12]. However, changes in diagnostic practices do not provide the full explanation for the increase in the rates of hospitalization with septicemia/sepsis in the US, particularly prior to 2009. Indeed, national trends in US hospitalization rates with septicemia and/or sepsis present on the diagnosis between 2003-2009 closely resemble the trends in hospitalization rates with both an infection and mechanical ventilation present on the diagnosis (Figure 1 in [13]). The increase in the rates of such hospitalizations associated with severe infection cannot be fully explained by changes in diagnostic practices. Therefore, factors behind the rise in the rates of hospitalization with septicemia/sepsis in the diagnosis, as well as the magnitude of those rates require a detailed examination.

There is good evidence that antimicrobial resistance increases mortality risk for sepsis, importantly by reducing the probability of initially appropriate antibiotic therapy (IIAT) [14,15]. However, less is known about the contribution of antibiotic use or the prevalence of antibiotic resistance to rates of hospitalization with septicemia/sepsis. Antibiotic resistance and use can contribute to the volume of hospitalizations associated with bacterial infections, particularly the more severe ones, through several mechanisms [16]. Antibiotic resistance can facilitate the progression of an infection into a more severe outcome due to failure of initial therapy to clear the infection. This mechanism would create an association between antibiotic resistance and more severe types of infection. For example, antibiotic resistance in Enterobacteriaceae, including fluoroquinolone resistance in *Escherichia coli* was found to be associated with a more severe presentation in urinary tract infections [17,18]. In England, prevalence of amoxicillin-clavulanate

(co-amoxiclav) resistance in *E. coli*-associated bloodstream infections is significantly higher than in *E. coli*-associated urinary tract infections (Figures 2.1 vs. 2.7 in [49]). For some pathogens, antibiotic use and resistance can also contribute to the overall increase in the incidence of bacterial infections that could lead to severe illness episodes. For example, receipt of fluoroquinolones was found to be a risk factor for a methicillin-resistant *Staphylococcus aureus* (MRSA) infection [19-22], while receipt of different antibiotics, particularly third generation cephalosporins was found to be a risk factor for subsequent colonization/infection with extended-spectrum beta-lactamase-producing *Klebsiella* species and *E. coli* [23-25]. Furthemore, reduction in ciprofloxacin prescribing in a London hospital was associated with a decline in the incidence of MRSA infections [26]. We also note that several biological mechanisms for the relation between fluoroquinolone use and colonization with fluoroquinolone-resistant staphylococci were proposed [27,28].

In a recent study (Goldstein et al. 2018, in submission), we have documented associations between state-specific prevalence of antibiotic resistance for several combinations of antibiotics/bacteria, and state-specific rates of hospitalization with septicemia in US adults. Among the 31 combinations of bacteria/antibiotics studied in (Goldstein et al. 2018, in submission), resistance to fluoroquinolones in *E. coli* had the strongest association with septicemia hospitalization rates in adults aged 50-64y, 65-74y, 75-84y, and 85+y. *E. coli* is a major source of Gram-negative septicemia in the US [1], and prevalence of fluoroquinolone resistance in *E. coli* isolates in both urinary tract and bloodstream infections in the US is high [29-32].

Changes in antibiotic use, both in the inpatient and outpatient settings may be accompanied by corresponding changes in rates of hospitalizations associated with bacterial infections, including sepsis. At the population level, a decline in fluoroquinolone and cephalosporin use in England was accompanied by a major drop in the rates of illness and death associated with *Clostridium difficile* and MRSA infections [33-36]. The volume of fluoroquinolone and cephalosporin prescribing in England declined between 2006-2013, both overall and in the inpatient setting (about 50% reduction) [33]. During the same period, incidence of *C. difficile* infections and MRSA infections and bacteremia in England, as well as the rates of associated deaths dropped by over 80% [33-36]. Additionally, the proportion of *C. difficile* infections resistant to fluoroquinolones declined drastically (e.g. a reduction from 67% in 2006 to 3% in 2013 in Oxfordshire, [33]). Rates of both community- and hospital-associated MRSA bacteremia in England also showed pronounced declines after 2006 (Figures 1 and 3 in [35]).

The US had more concerning trends in both *C. difficile* infection rates and MRSA infection rates during the period between 2006-2012 than the UK. In contrast to sharp declines in the UK, *C. difficile* infections in the US continued to increase [37]. While the rates of invasive MRSA infection in the US declined between 2005-2011 [38], the drop in the rates of MRSA bacteremia in England during the same period was sharper, particularly for community-associated bacteremia (compare Figures 1 and 3 in [35] with [38]). One factor in these contrasting patterns may be contrasting patterns of fluoroquinolone use. In the US, fluoroquinolone use in adults more than doubled between 1995-2002 [39], then declined in the hospital setting [40]. At the same time, fluoroquinolone outpatient prescription rates to adults were stable between 2000-2010 [41], and stable in older adults between 2011-2014 [42]. The discrepancy in the trends in outpatient fluoroquinolone prescribing in the US vs. UK may partly explain why the rates of community-associated invasive MRSA infections in the US remained flat between 2005-2011 [38], while rates of community-acquired MRSA bacteremia in the UK dropped by more than half after 2006 (Figures 1 and 3 in [35]). We also note the high prevalence of fluoroquinolone (levofloxacin) non-susceptibility in community-associated MRSA isolates in the US in recent years [43].

A recent US study reported a 17.5% decrease in the incidence rates of long-term-care onset *C. difficile* infection in 10 US sites between 2011-2015, concomitant with the ongoing decrease in inpatient fluoroquinolone use [44]. This is an encouraging reversal of the earlier trends [37], supporting the value of further reductions in fluoroquinolone prescribing. In recent years, FDA has issued recommendations against fluoroquinolone use for certain conditions (e.g. uncomplicated urinary tract infection) due to potential adverse effects [45]. It remains to be seen what the effect of those recommendations will be on fluoroquinolone prescribing in the US, particularly in the outpatient setting, and the rates of severe outcomes to which fluoroquinolone use and resistance contributes.

Reduction in fluoroquinolone use in England was not associated with a reduction in the rates of severe outcomes stemming from all bacterial pathogens for which fluoroquinolone resistance prevalence was previously high. Levels of *E. coli* and *Klebsiella*-associated bacteremia were continuing to rise in England after 2006 [46,47]. Amoxicillin-clavulanate (co-amoxiclav) is used as empiric treatment for many infection syndromes in the UK, particularly in the inpatient setting [48], and its use in the England increased significantly after 2006 [49]. Moreover, incidence of bacteremia with *E. coli* strains resistant to amoxicillin-clavulanate began to increase rapidly after 2006 ([47], Figure 4), and this rise was faster than the rise in incidence of bacteremia with susceptible strains [47]. By 2014, over 40% of *E. coli*-associated bloodstream infections in England were amoxicillin-clavulanate-resistant (Figure 2.1 in [50]). This suggests that use of and resistance to different antibiotics may help sustain the growth in the rates of *E. coli*-associated sepsis, and that patterns in resistance to various antibiotics in *E. coli* should be accounted for when antibiotic replacement is being considered. We note that amoxicillin-clavulanate use in England began to decrease after 2012, and trends in amoxicillin-clavulanate resistance among E. coli-associated urinary tract infections (UTIs) began to change as well ([50], Figures 2.1, 3.4). In the US, *E. coli* isolates from individuals hospitalized for urinary tract infections (UTIs) and bloodstream infections have high levels of amoxicillin and multidrug resistance [31,32], while most UTI *E. coli* samples resistant to third-generation cephalosporins in hospitalized patients in [30] were also resistant to fluoroquinolones. While the results in (Goldstein et al. 2018, in submission) support the contribution of fluoroquinolone resistance prevalence to the rates of *E. coli*-associated septicemia, options for reducing the incidence of *E. coli*-associated sepsis in the US through replacement of fluoroquinolones by other antibiotics may be limited. For urinatry tract infections, some of the options may potentially be provided by antibiotics like nitrofurantoin and fosfomycin that are largely reserved for UTIs and may be less likely to generate resistance compared to antibiotics that are prescribed more widely for a variety of conditions. Data from Veterans Affairs hospitals in the US suggest low prevalence of non-susceptibility to nitrofurantoin in *E. coli*-associated UTI hospitalizations [31]. Data from the English Surveillance Programme for Antimicrobial Utilisation and Resistance (ESPAUR) suggest low prevalence of non-susceptibility to nitrofurantoin and fosfomycin in *E. coli*-associated UTIs, both for the community and acute hospital isolates (Figure 2.7 in [50]). Data on resistance patterns to different antibiotics for different pathogens that cause various syndromes (including UTI-related syndromes) are needed to inform the corresponding treatment guidelines.

The results in (Goldstein et al. 2018, in submission) suggest that prevalence of resistance to commonly prescribed antibiotics contributes to the rates of hospitalization with septicemia/sepsis in the US. Those findings support the need for enhancing antibiotic stewardship, including for fluoroquinolone use, both in the inpatient and outpatient settings [51], and for preventing acquisition of antibiotic-resistant bacteria (particularly *E. coli* in the elderly, e.g. [52,53]). The findings in (Goldstein et al. 2018, in submission) about the association of fluoroquinolone resistance in *E. coli* with the rates of hospitalization with septicemia complement the evidence from the UK about the parallel declines of fluoroquinolone resistance/use and the rates of severe outcomes associated with bacterial infections, including *C. difficile* infections [33-35] and MRSA infections and bacteremia [34-36,26]. All this evidence points to the potential benefit of reducing fluoroquinolone use in the US, both in the outpatient and the inpatient settings. A recent study found that about 25% of outpatient fluoroquinolone prescriptions to US adults in 2014 were for conditions that did not require antibiotics, or where fluoroquinolones are not recommended first-line therapy [54]. These considerations might reflect the tendency to use commonly prescribed antibiotics for a variety of conditions, including those for which more narrow-spectrum antibiotics are available and recommended. For *E. coli*-associated urinary tract infections, data from both the US and the UK suggest that prevelance of resistance to some of the antibiotics that are mostly recommended for UTI treatment, such as nitrofurantoin and fosfomycin is significantly lower than prevalence of resistance to the more widely prescribed antibiotics, including fluoroquinolones, ampicillin, amoxicillin, amoxicillin-clavulanate, and trimethoprim/sulfamethoxazole [31,50]. The UK experience in the reduction of fluoroquinolone and cephalosporin use also suggests that for some pathogens, particularly *E. coli,* resistance to some of the replacing antimicrobials (e.g. amoxicillin-clavulanate) may contribute to the rates of associated bacteremia hospitalizations [47]. Taken together, these considerations suggest that treatment guidelines for each particular syndrome should account for patterns of resistance to different antibiotics for various pathogens that are causative agents for the given syndrome [36,47]. The study of patterns of antibiotic resistance requires surveillance data on the prevalence of disease and resistance associated with different pathogens [36] akin to the one described in [34-36,46-48,50]. We note that such surveillance data are less readily available in the US, particularly for sepsis and bacteremia hospitalizations. Additionally, evidence on the causal effect between patterns in antibiotic resistance and use and the incidence of hospitalizations associated with bacterial infections is limited. Further studies in that regard examining antibiotic receipt prior to bacteremia/sepsis hospitalization and diagnosis (e.g. during treatment of UTIs, gastrointestinal infections and other syndromes) should help inform public health policies. Finally, the high burden of severe outcomes stemming from bacterial infections that is associated with antibiotic resistance and use, and the limited options for antibiotic replacement support the need for a robust supply of new antibiotics for common prescription. This need may not be fulfilled through the current system of antibiotic development/production, and additional incentives for antibacterial research and development are worth considering [55].